\documentclass[pra,showpacs,twocolumn,superscriptaddress,amsmath,amssymb]{revtex4-1}

\usepackage{graphicx}

\usepackage{color}

\usepackage[colorlinks=true,linkcolor=red,citecolor=blue,urlcolor=blue]{hyperref}

\begin{document}

\title{Prospects for assembling ultracold radioactive molecules from laser-cooled atoms}

\author{Jacek K{\l}os}
\affiliation{Department of Physics, Temple University, Philadelphia, Pennsylvania 19122, USA}
\affiliation{Joint Quantum Institute, National Institute of Standards and Technology and University of 
Maryland, Gaithersburg, Maryland 20899, USA}
\author{Hui Li}
\affiliation{Department of Physics, Temple University, Philadelphia, Pennsylvania 19122, USA}
\author{Eite Tiesinga}
\affiliation{Joint Quantum Institute, National Institute of Standards and Technology and University of 
Maryland, Gaithersburg, Maryland 20899, USA}
\author{Svetlana Kotochigova}
\email{skotoch@temple.edu}
\affiliation{Department of Physics, Temple University, Philadelphia, Pennsylvania 19122, USA}

\date{\today}

\begin{abstract}
Molecules with unstable isotopes often contain heavy and deformed nuclei and thus possess a
high sensitivity to  parity-violating effects, such as Schiff moments. Currently the best limits on Schiff moments 
are set with diamagnetic atoms. Polar molecules with quantum-enhanced sensing capabilities, however, can offer better 
sensitivity. In this work, we consider the prototypical $^{223}$Fr$^{107}$Ag molecule, as the octupole deformation of 
the unstable $^{223}$Fr francium nucleus amplifies the nuclear Schiff moment of the molecule by two orders of 
magnitude relative to that of spherical nuclei and as the silver atom has a  large electronegativity. 
To develop a competitive experimental platform based on molecular quantum systems,  $^{223}$Fr atoms and  
$^{107}$Ag atoms have to be brought together at ultracold temperatures. 
That is, we explore the prospects of  forming  $^{223}$Fr$^{107}$Ag from laser-cooled Fr and Ag atoms.  
We have performed fully relativistic electronic-structure calculations of ground and excited states of FrAg that account for 
the strong spin-dependent relativistic effects of Fr and the strong ionic bond to Ag. 
In addition, we predict the nearest-neighbor densities of magnetic-field Feshbach resonances in ultracold 
$^{223}$Fr+$^{107}$Ag collisions with coupled-channel 
calculations. These resonances can be used for magneto-association into ultracold, weakly-bound FrAg. We also 
determine the conditions for creating $^{223}$Fr$^{107}$Ag molecules in their absolute ground state  
from these weakly-bound dimers via stimulated Raman adiabatic passage using our calculations of the relativistic 
transition electronic dipole moments.
\end{abstract}

\maketitle


\section{Introduction}

Engineered quantum matter holds  promise for quantum computation as well as
the development of novel materials and sensors. Quantum technologies based on atoms and
atomic ions have partly fulfilled  these promises
\cite{Cirac2004,Weiss2017,Ludlow2015,Brewer2019}. Ultracold, sub-millikelvin molecules
represent a next frontier for controlling quantum matter. The richness of their internal
states has established molecules as promising precision-measurement tools in quantum science,
assisted by long coherence times among internal states in laser-based optical traps
\cite{DeMille2002,Ni2018,Hudson2018}. Key to these advances has been the development of
cooling techniques, which sufficiently reduce the entropy of the molecules in order to apply
ever more refined quantum control techniques.

Advancing fundamental physics and related precision measurements often require the creation of
unexplored polar molecules \cite{Hinds1997}. Current efforts in this direction focus on
sensors of  fundamental interactions and forces using cold and
ultracold molecules \cite{Safronova2018,Hutzler2020}. Here, the search for  forces that violate both
time-reversal (T) and parity (P) invariance is of fundamental importance for physics beyond
the standard model.  P,T-odd nuclear interactions, for example, give rise to the nuclear
Schiff moment \cite{Sushkov1984}, which may interact with the electrons in the molecule and
lead to measurable shifts in molecular spectra \cite{Skripnikov2020}.

\begin{figure}
\includegraphics[angle=0,scale=0.6, trim=100 250 360 150,clip]{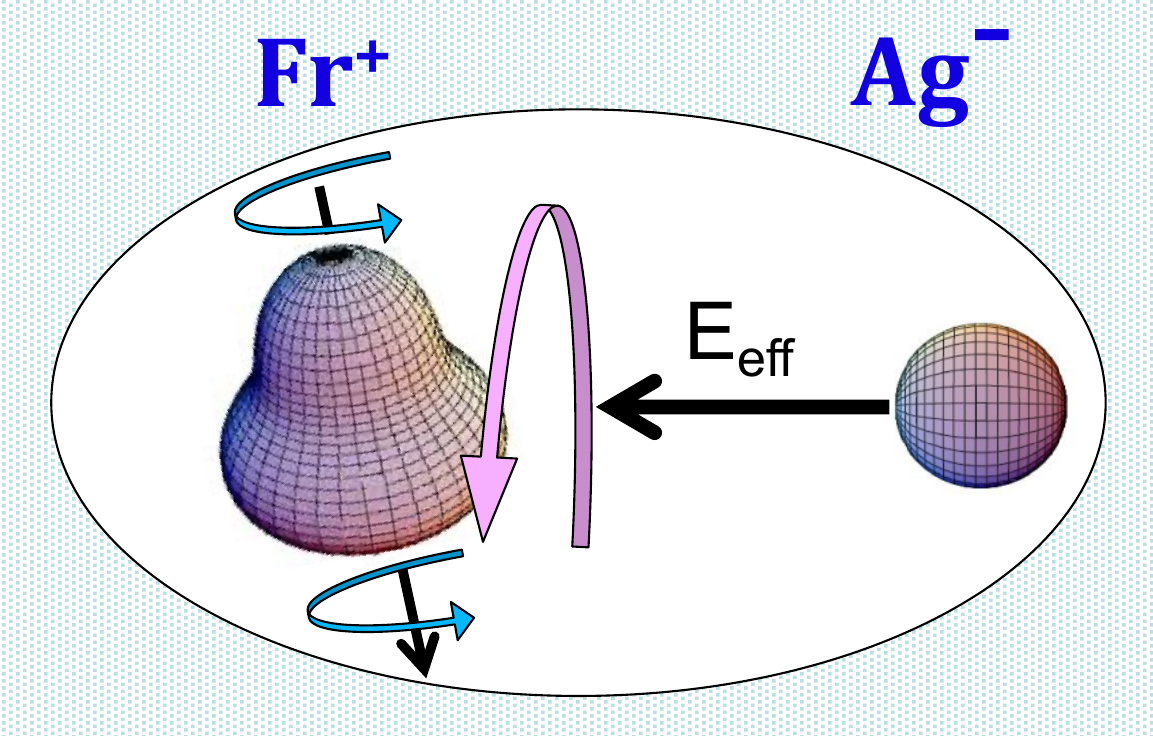}
\caption{Beyond-standard-model quantum sensor based on FrAg molecules containing unstable Fr with a deformed 
non-spherical nucleus. The strong ionic bond of FrAg leads to a strong internal electric field ${\cal E}_{\rm eff}$
significantly enhancing the sensitivity to  parity-violating effects.}
\label{fig:FrAg}
\end{figure}

Among the candidates for  measurements of the nuclear Schiff moment are molecules containing
unstable isotopes of radium (Ra) and francium (Fr), which have an octopole-deformed nucleus and
thus possess a high sensitivity to  parity violation \cite{Isaev2010,Isaev2017,Kudashov2014,
Berger2020,Hutzler2021,Fleig2021}.  Both heavy atom species are now routinely cooled and
trapped in magneto-optical traps  despite  their short lifetime by either $\alpha$ or $\beta$ decay \cite{Gomez2007,
Tandecki2013}. In the search for bonding partners for Ra and Fr two 
criteria must be considered: {\it i}) bonding partners must  have a large electron affinity that leads
to an ionic bond and a strongly polarized Ra or Fr atom; {\it ii})  being amenable to laser
cooling and trapping. An ionic bond is also correlated with a large permanent
molecular electronic dipole moment and with a large effective electric field, ${\cal E}_{\rm eff}$,
acting on either the unstable nucleus or the electrons \cite{Flambaum1985, Hinds1997,
Fleig2021}. Among the most relevant partner for both Ra and Fr is  the silver (Ag) atom
\cite{Fleig2021, DeMille2021}. It satisfies both criteria as having a large electronegativity
of $hc\times10\,521$ cm$^{-1}$ \cite{Bilodeau1998}, and having been laser cooled
\cite{Uhlenberg2000}. Here, $h$ is the Planck constant and $c$ is the speed of light in
vacuum.

In this paper, we consider the prototypical francium-silver molecule FrAg, shown in Figure~\ref{fig:FrAg} for the
development of a quantum sensor in search of a nuclear Schiff moment.  The idea is to assemble
FrAg molecules from laser-cooled $^{223}$Fr and $^{107}$Ag atoms \cite{DeMille2021}. Both
atoms have an electron-spin-1/2 or $^2$S electronic ground state, while their electronic
molecular ground state is well described as an electron-spin-zero or singlet $^1\Sigma^+$
Hund's case (a) state \cite{Herzberg,Nikitin1994},  similar to that for  the ground state of
bi-alkali-metal molecules.  Alkali-metal dimers have already been assembled
from ultra-cold atoms and been shown to be scientifically relevant \cite{Ni08,Ye2017}.

We first assume that the ultracold atoms are prepared in their energetically lowest Zeeman,
hyperfine state and collide in the presence of an external magnetic field and from there can be  
bound together with a small binding energy of order $hc\times 10^{-3}$ cm$^{-1}$,
in an electronic configuration that is predominantly of triplet a$^3\Sigma^+$ character.  This
binding process is either achieved via a slow time-dependent sweep or ramp of the magnetic 
field near a Fano-Feshbach resonance, also known as magneto-association, or via  microwave radiation
near such resonances \cite{Chin2009}. We will show that useable Feshbach resonances exist
in  $^{223}$Fr and $^{107}$Ag collisions.

The next step is to search for a route, based on stimulated Raman adiabatic
passage (STIRAP)  processes \cite{STIRAP},  to coherently transfer 
the population from a weakly-bound rovibrational state to the strongly-bound rovibrational ground 
state of the X$^1\Sigma^+$ state via  a rovibrational state of the mixed and coupled b$^3\Pi$ and A$^1\Sigma^+$ excited 
electronic  states. Mixing is due to relativistic spin-orbit interactions, which requires us to label
electronic state with the 
Hund’s case (c) coupling scheme rather than the case(a) scheme so far\cite{Herzberg,Nikitin1994}.

The two-step formation of ultracold FrAg molecules from ultracold Fr and Ag is made
challenging due to a  lack of knowledge of their relativistic electronic, rovibrational, and
hyperfine structure in both electronic ground and excited states.  To our knowledge,  only the electronic singlet 
and triplet ground-state potentials of FrAg have been calculated \cite{Tomza2021}.  Here, we  describe our 
theoretical study of potentials, electric dipole moments, and rovibrational states of FrAg. In addition, this includes
the prediction of Feshbach resonance densities and locations, as well as the development of 
Raman schemes for the formation of the absolute ground state of FrAg. Unless otherwise noted,
we present results for  rovibrational states of the $^{223}$Fr$^{107}$Ag isotopologue.

\section{Results}

\subsection {Electronic potentials and transition dipole moments}

We begin with the determination of the adiabatic potential energy surfaces of FrAg electronic states
as well as transition electronic dipole moments between these states as functions of
atom-atom separation $R$.  
Here, relativistic electronic structure calculations using the DIRAC computational suite
\cite{Dirac21} enable us to account for spin-orbit effects on  FrAg states.
This includes spin-orbit coupling between the A$^1\Sigma^+$ and b$^3\Pi$ states as well as  the 
weaker second-order spin-orbit splitting of the a$^3\Sigma^+$ state.

Adiabatic potentials are uniquely labeled by $n(\Omega^\sigma)$ within the
Hund’s case (c) notation, where $\Omega$ is the absolute value of the projection of the total
electronic angular momentum on the internuclear axis and $\sigma=\pm$ represents the even or
odd reflection symmetry of the electron wavefunction through a plane containing the
internuclear axis when $\Omega=0$. Finally, $n=1,2,\dots$ labels states of the same
$\Omega^\sigma$ value ordered by increasing energy. Then, the energetically-lowest
$n(\Omega^\sigma)=1(0^+)$  state connects to the Hund's-case-(a) X$^1\Sigma^+$ state  while
the a$^3\Sigma^+$ state has $1(0^-)$ and $1(1)$ components.  The A$^1\Sigma^+$ and b$^3\Pi$
states mix to form $\Omega^\sigma=0^+$, $0^-$, 1, and 2 states, but in this article we will
mostly be interested in the $2(0^+)$ and $3(0^+)$ states.  We also determine the $R$-dependent
transition dipole moments between $n(\Omega^\sigma)$ ground and excited states.  
A description of  electron orbitals used in the DIRAC calculations and values for the long-range van-der-Waals 
and other dispersion coefficients can be found in Appendix \ref{App1}.

\begin{figure}
\includegraphics[scale=0.3,trim=0 0 0 0,clip]{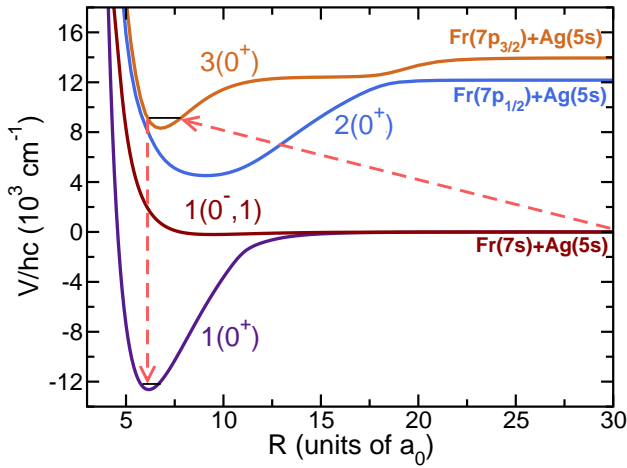}
\caption{Relevant electronic potentials of FrAg as functions of atomic separation $R$.
Potentials are identified by Hund's case (c) state labels and for large
separations by atomic states. The zero of energy 
is at the dissociation limit or threshold of the $1(0^+)$ and $1(0^-, 1)$ states. 
The small splitting between the $1(0^-)$ and $1(1)$ states is invisible on the scale of this figure.
Short black lines in the $1(0^+)$ and $3(0^+)$  potentials indicate rovibrational levels of these potentials.
The dashed  lines with arrows connecting these levels with near threshold bound states
 indicate a possible STIRAP  pathway to the FrAg rovibrational 
ground state.}
\label{fig:PESs}
\end{figure}

\begin{table}
\caption{Spectroscopic constants of   $^{223}$Fr$^{107}$Ag  in  relativistic ground and excited states relevant to the STIRAP scheme. These include equilibrium interatomic separation $R_{\rm e}$, dissociation energy $D_{\rm e}$, harmonic spring constant $k$, harmonic (angular) frequency $\omega_{\rm e}$, and rotational constant $B_{\rm e}$. Data is compared
to the non-relativistic results of Ref.~\cite{Tomza2021}  where available.}
\label{tb:rel_pot_params}
\begin{tabular}[t]{|c|c|c|c|c|c|}
\hline
\hline
        State        &       $R_{\rm e}/a_0$      &      $D_{\rm e}/hc$  & $k/hc$  & $\omega_{\rm e}/2\pi c$  & $B_{\rm e}/hc$   \\
                       &         &    (cm$^{-1}$) &  (cm$^{-1}/a_0^2$) &  (cm$^{-1}$)  &  (cm$^{-1}$) \\
\hline    
1(0$^+$) & 6.164 &12635 &4391.5&85.54&0.0219\\
Non-rel.~\cite{Tomza2021}&6.190 &12700 &-&84.2&0.0215\\
\hline
1($0^-$) &9.422 &205&68.942&10.72&0.0094\\
Non-rel.~\cite{Tomza2021} & 9.451& 193&-&10.6&0.0093\\ 
\hline
2(0$^+$)&9.100&4017&342.18&23.88&0.0010\\
\hline
3(0$^+$)&6.740&8125&2093.2 &59.06&0.0183\\
\hline
\end{tabular}
\end{table}

Figure~\ref{fig:PESs} shows our results for electronic relativistic adiabatic potentials $V_{n(\Omega^\pm)}(R)$ relevant for transferring
population from  weakly-bound Feshbach molecular states to the absolute rovibrational ground
state of the 1$(0^+)$ potential. In the figure atom separations are expressed in units of
$a_0=0.05292$ nm, the Bohr radius. Spectroscopic constants for these potentials can be found in
Table \ref{tb:rel_pot_params}, while a graph with additional excited electronic potentials can be found in
Appendix~\ref{App1}. 
The dissociation energy of  our $1(0^+)$ potential is only 0.5\,\% smaller than that of Ref.~\cite{Tomza2021}  based on
non-relativistic calculations. The corresponding fractional difference for the shallow $1(0^-)$ potential is about 10\%.
The absolute difference, however, is only $hc\times 12$ cm$^{-1}$.
There is also noticeable difference in their harmonic constant $\omega_e$, $hc\times85.5$ cm$^{-1}$ for relativistic potential versus $hc\times84.2$ cm$^{-1}$ for non-relativistic potential. 

At first glance, there is similarity with the potential surfaces for heavy di-atomic alkali-metal molecules. This is due 
to the single active open valence orbital of alkali-metal and silver atoms. Thus we find a deep $1(0^+)$ ground state 
and  shallow nearly-degenerate $1(0^-)$ and  $1(1)$ excited states that correlate to the non-relativistic Hund's case (a)-like X$^1\Sigma^+$ 
and a$^3\Sigma^+$ states, respectively. These states dissociate to two  $^2$S ground-state atoms.  
Next, we observe the avoided crossings between the $2(0^+)$ and $3(0^+)$ levels that dissociate
to an excited $^2{\rm P}_{j=1/2}$ or $^2{\rm P}_{j=3/2}$ Fr atom and a ground-state Ag atom. For alkali-metal dimers these 
states also exist. The two states are the result of spin-orbit mixing of the non-relativistic A$^1\Sigma^+$ and b$^3\Pi$ states.
Near avoided crossings the potentials for these non-relativistic states cross.

There  are significant differences between the potentials of FrAg and alkali-metal dimers as well.  First,  the ground 
$1(0^+)$ potential for FrAg is more than twice as deep at its equilibrium separation as the corresponding potential of 
KRb  \cite{Ni08}, RbCs \cite{Tiemann2011,RbCs2012}, and Cs$_2$ \cite{Sainis2012}. 
The bond in FrAg is far more ionic. On the other hand, the depths of
the shallow $1(0^-,1)$ states are very similar. 
Second, the shape of the excited  $2(0^+)$ and $3(0^+)$ potentials differ in two important ways. The
extended, flat minimum of the $2(0^+)$ state between $R=8a_0$ and $11a_0$ is seen to avoid with
the $1(0^+)$ state. In alkali-metal dimers, the harmonic (spring) constant near the
equivalent minimum of the $2(0^+)$ state is significantly larger and the avoided crossing with
the X$^1\Sigma^+$ potential much less pronounced.  Finally, for FrAg the $2(0^+)$ and $3(0^+)$
potentials have an avoided crossing on their inner walls, where the slope of the potentials
with respect to $R$ is negative. For alkali-metal dimers this avoided crossing occurs for
separations, where the slope of $2(0^+)$ potential is already positive.

\begin{figure} 
\includegraphics[scale=0.3,trim=0 0 0 0,clip]{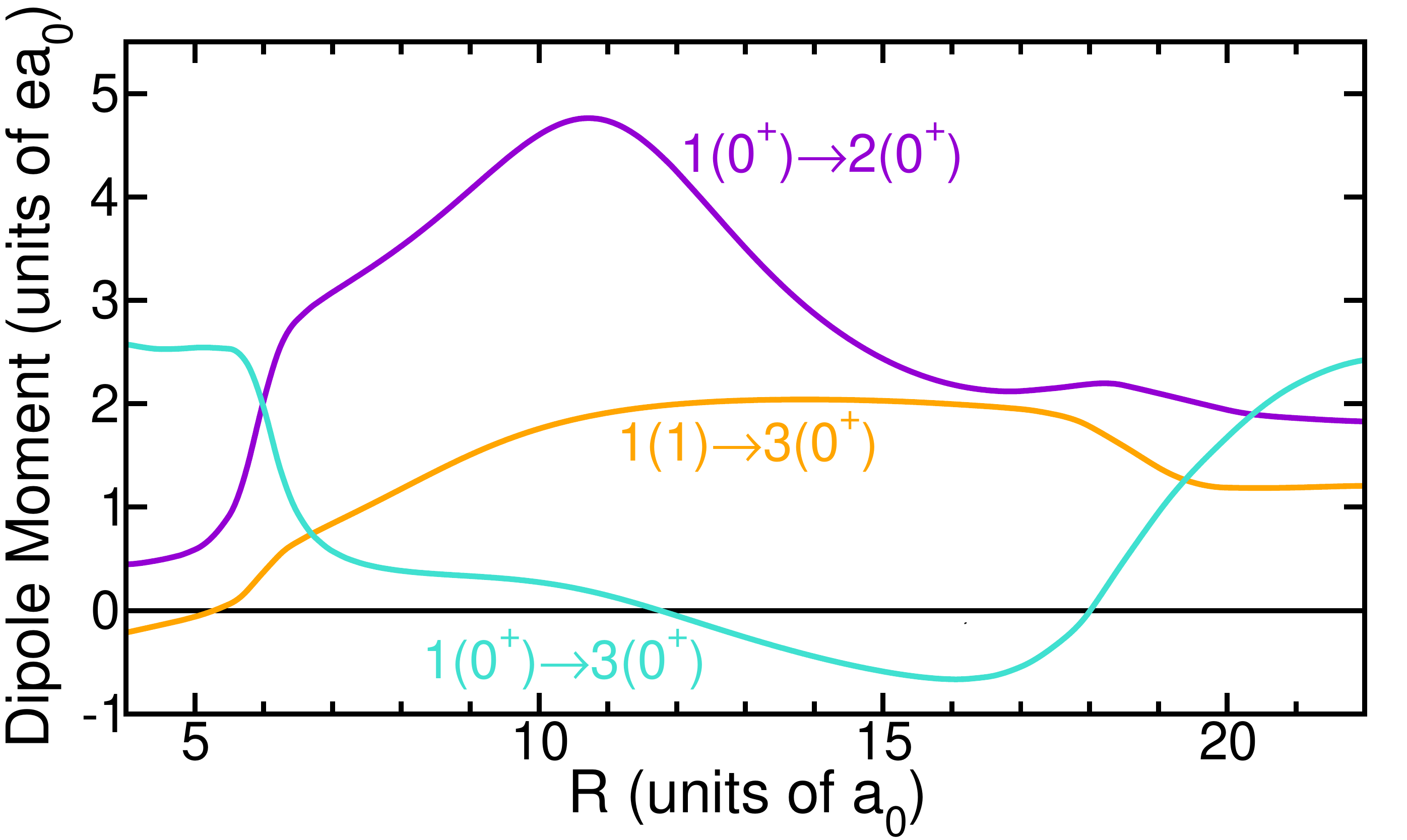}
	\caption{Relevant electronic transition dipole moments $d$ of FrAg  as functions of separation $R$.   The moments have been found from relativistic calculations. The purple, turquiose, and orange curves are for the $1(0^+)\to 2(0^+)$, $1(0^+)\to 3(0^+)$, and $1(1)\to 3(0^+)$  transitions, respectively. }
\label{fig:TDMs}
\end{figure}

We have also determined  electronic transition dipole moments between ground and excited electronic states.  Computational 
details can be found in Appendix~\ref{App1}. Three of these dipole moments as functions of internuclear separation $R$ 
are shown in Fig.~\ref{fig:TDMs}.  First, we observe that dipole moments undergo rapid changes near $6a_0$ and 
$19a_0$ corresponding to the avoided crossings between the $2(0^+)$ and $3(0^+)$ potentials in Fig.~\ref{fig:PESs}.
Second, we will mostly pay attention to the two transition dipole moments to the $3(0^+)$ state. 
The dipole moments are large, of order $ea_0$, and are non-zero for most $R$. Here, $e$ is the elementary charge.
Especially noticeable is the large dipole moment at the asymptotic limit due to spin-orbit mixing of the 7p$_{1/2}$ and 7p$_{3/2}$ 
excited levels of Fr.  As we will show, this promises efficient transfer from a weakly-bound $1(1)$ molecule to the 
strongly-bound $1(0^+)$ molecule.

\begin{figure}
\includegraphics[scale=0.3, trim=0 0 0 0,clip]{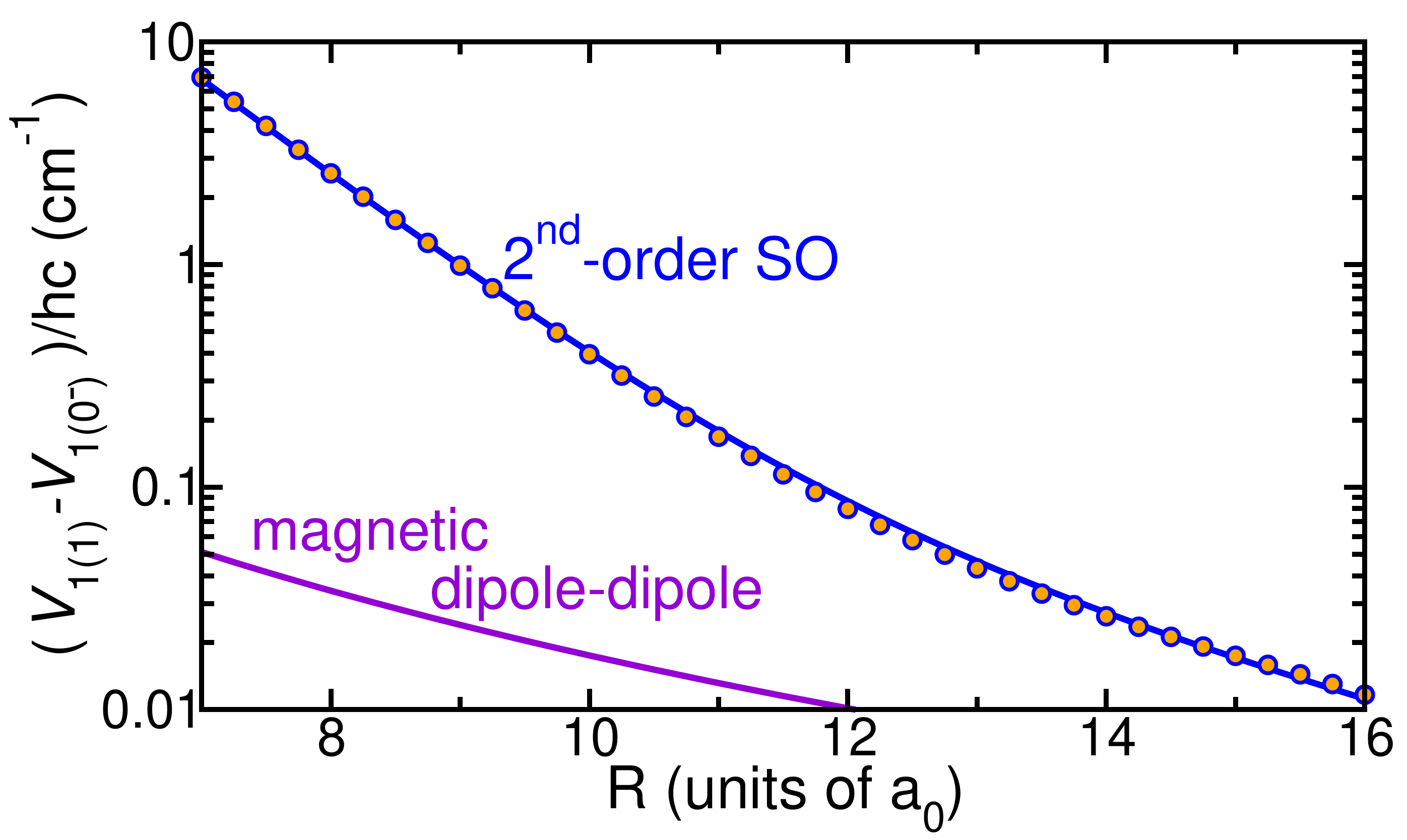}
\caption{The second-order spin-orbit splitting (filled blue circles) of FrAg, defined as the
potential energy of the $\Omega=1$ component minus that of the $\Omega=0^-$ component of the
a$^3\Sigma^+$ state, as a function of interatomic separation $R$  based on relativistic
electronic-structure calculations.  The blue curve is a fit to this data using the functional
form given in the text. The purple curve shows {\it minus} the corresponding splitting from the
magnetic dipole-dipole interaction due to the magnetic moments of the electrons.
}
\label{fig:2ndSO}
\end{figure}

For a precise description of ultra-cold Fr and Ag collisions, we need
the splitting between the $1(0^-)$ and $1(1)$ components of the a$^3\Sigma^+$ potential. This is the
second-order spin-orbit interaction and is shown in Fig.~\ref{fig:2ndSO}.  Its behavior is determined by the overlap of electron
wavefunctions from each atom and, thus, decreases exponentially with increasing 
separation $R$.  For later use, the data has been fit to 
\begin{equation}
V_{1(1)}(R) -V_{1(0^-)}(R) = A_1e^{-B_1(R-R_1)} + A_2e^{-B_2(R-R_2)},
\end{equation}
where $A_1/hc=2.358\,24$ cm$^{-1}$, $B_1= 1.017\,01 a^{-1}_0$, $R_1=8a_0$
and  $A_2/hc=0.022$ cm$^{-1}$, $B_2= 0.37 a^{-1}_0$, $R_2=14a_0$.
For later use we  define the singlet X$^1\Sigma^+$ potential $V_{\rm X}(R)\equiv V_{1(0^+)}(R)$
and triplet a$^3\Sigma^+$ potential  $V_{\rm a}(R)=(V_{1(0^-)}(R)+2V_{1(1)}(R))/3$ 
(See also App.~\ref{App1}.)

A comparison of the $2^{\rm nd}$-order spin-orbit interaction of FrAg with that of heavy
alkali-metal dimers, such as RbCs \cite{RbCs2012}, shows that the former is almost ten
times stronger at the inner-turning point of the a$^3\Sigma^+$ potential near $8a_0$ when the 
potential energy is that of the dissociation limit or atom-atom threshold.
In Fig.~\ref{fig:2ndSO}, we also show {\it minus} the splitting between $1(1)$ and $1(0^-)$ due to
the magnetic dipole-dipole interaction between the magnetic moments of electron spins of Fr
and Ag. This dipole-dipole interaction is of order $E_{\rm h}a_0^3\alpha^2/R^3$, where $E_{\rm h}$ is the 
Hartree energy and $\alpha$ is the fine-structure constant. It is small for the separations shown in the figure, but 
will dominate for $R>20a_0$.

\subsection {Magnetic Feshbach resonances in ultracold Fr+Ag collisions}\label{sec:fb}

We can now  describe results for  ultra-cold, $\mu$K collisions of $^{223}$Fr and $^{107}\!$Ag in their $^2$S 
electronic ground state as well as the near-threshold, weakly-bound ro-vibrational states of  $^{223}$Fr$^{107}\!$Ag.
Specifically, we describe collisional magnetic Feshbach resonances when these atoms are prepared in their 
energetically-lowest electronic, hyperfine, and  Zeeman states in the presence of an external magnetic field
with strength $B$. 
These resonances are due to mixing of the R-dependent molecular interactions by the
 Zeeman and hyperfine or Fermi-contact interactions of the $^2$S atoms.

The Hamiltonian $H$ for the relative motion of  $^{223}$Fr and $^{107}\!$Ag is similar to that of interacting ground-state 
hydrogen atoms or alkali-metal atoms. Following Ref.~\cite{Stoof1988}, the  atoms are assumed to be
point-like with a mass equal to that of the atoms. 
Each atom has an electron spin (quantum number) equal to $1/2$ and  a non-zero
nuclear spin, whose value is unique to the actual isotope.  Here,  $^{223}$Fr has nuclear spin 3/2 and $^{107}\!$Ag 
has nuclear spin 1/2.
Electron and nuclear spin of each atom  are coupled by the Fermi-contact and Zeeman interactions.
Relevant hyperfine constants, $g$ factors, and masses, are taken from Refs.~\cite{Dahmen1967,Coc1985,Ekstrom1986,Stone2014,Huang2021},
where  $^{107}\!$Ag has an ``inverted'' hyperfine structure. The Fermi-contact  coefficient of $^{223}$Fr is many 
times larger than the absolute value of that of  $^{107}\!$Ag.

\begin{figure}
	\includegraphics[scale=0.31, trim=0 0 0 0,clip]{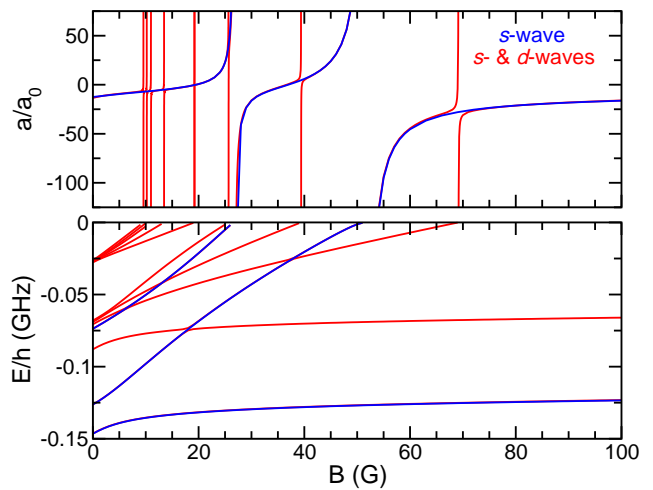}
\caption{Scattering length, $a$, (top panel) for colliding $^{223}{\rm Fr}$ and $^{107}\!$Ag  atoms
 in their energetically-lowest  hyperfine state and their near-threshold bound-state energies, $E$,
(bottom panel) as functions of magnetic field $B$ up to
100~G. Blue and red curves correspond to calculations including channels with only $\ell=0$ ($s$-) and 
$\ell=0,2$ ($s$-, $d$-) partial waves, respectively.  
}
\label{fig:FBsmallB} 
\end{figure}

The Hamiltonian also contains the relative kinetic energy operator, which is an operator in the separation
between the two atoms $R$ and the orientation of their interatomic axis $\hat R$. Eigenfunctions 
of the orientation-dependent part of the kinetic energy operator are spherical harmonic functions in $\hat R$ 
labeled by orbital angular momentum or partial wave $\ell$ and its projection $m_\ell$  along the magnetic field direction.
In addition, $H$ includes isotropic molecular interactions that only depend on separation $R$. The 
isotropic potential for total molecular electron spin zero is $V_{\rm X}(R)$,  while that for total electron spin one
is $V_{\rm a}(R)$ as defined in the previous subsection. Finally, the Hamiltonian contains the weak $2^{\rm nd}$-order spin-orbit and 
magnetic dipole-dipole interactions. They are anisotropic, depend on the orientation of the total electron spin relative to  $\hat R$, 
and lift the $1(0^-)$ and $1(1)$ degeneracy. These weaker interactions mix molecular states 
with even $\ell$ (that is the $s$, $d$,\dots partial waves for $\ell=0,2,\dots$) or odd $\ell$ (that is the $p$, $f$,\dots partial waves). 

\begin{figure}
	\includegraphics[scale=0.31, trim=0 0 0 0,clip]{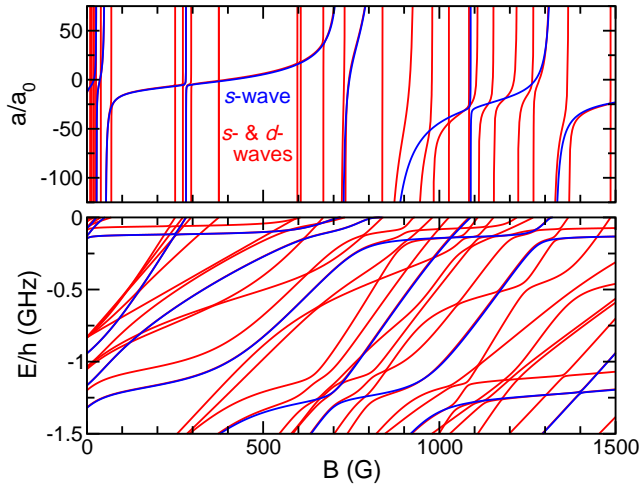}
\caption{Scattering length, $a$, (top panel) and near-threshhold bound-state energies, $E$,
(bottom panel) of  $^{223}{\rm Fr}+^{107}\!$Ag  as functions of magnetic field $B$ up
to 1500~G.  Other conditions, parameters, and definitions of line styles are as in
Fig.~\ref{fig:FBsmallB}.  }
	\label{fig:FBlargeB} 
\end{figure}

We have computed  the $s$-wave scattering length, $a$, as a function of magnetic field, a  
Feshbach resonance spectrum \cite{Chin2009},  for  ultracold $^{223}$Fr+$^{107}\!$Ag 
collisions. For the calculations, we rely on the coupled-channels method using our potential energy surfaces. 
The scattering length is determined from elastic $s$-wave scattering amplitudes at a collision energy of $k\times 1$ $\mu$K with entrance channel where $^{223}$Fr and $^{107}\!$Ag are  in their energetically-lowest
${m_{\rm Fr}=+1}$ and ${m_{\rm Ag}=-1}$ hyperfine state, respectively.
Here, $m_{X}$ with ${X={\rm Fr}}$ and Ag are  projections of the atomic angular momentum along the $B$-field
direction and $k$ is the Boltzmann constant.
For the calculations, allowed molecular coupled channels have conserved projection ${M_{\rm tot}=m_{\rm Fr}+m_{\rm Ag}+m_\ell=0}$ along the 
$B$-field direction and even values of $\ell$. For $^{223}$Fr+$^{107}\!$Ag with $M_{\rm tot}=0$, there are eight $\ell=0$ 
channels and thirty $\ell=2$ channels.
In addition, we have determined the Zeeman, hyperfine, rotation and vibration resolved 
near-threshold bound states with ${M_{\rm tot}=0}$. 

Figures~\ref{fig:FBsmallB} and \ref{fig:FBlargeB} show our computed scattering lengths $a(B)$ and threshold bound-state 
energies $E(B)$ relative to the entrance channel energy as function of magnetic field strength $B$ up to 1500 G. 
Here, 1 G equals 0.1 mT.  The figures show results from calculations that include only ${\ell=0}$  channels as well as those 
that include all ${\ell=0}$ and 2 channels. In both cases the scattering length has resonances, where its value rapidly 
goes through $\pm\infty$ with $B$.
The values for $a$ are mostly identical away from resonances for the two cases. In fact, the positions and (magnetic) widths
of those resonances found in both $s$-wave channel  and  $s,\,d$-wave channel calculations agree to a fraction of a Gauss. 
These resonances are $s$-wave Feshbach resonances, while the remaining resonances are $d$-wave  resonances.
Between 0~G and 1500~G, we find  seven $s$-wave   and  just over 30 $d$-wave Feshbach resonances.

Our analysis also  implies that the anisotropic interactions,  coupling $s$- and $d$-wave channels, are weak.
With some exceptions, the magnetic widths of $d$-wave resonances are smaller or narrower than those for $s$-wave
resonances. Adding larger partial-wave channels, that is $\ell=4,6,\cdots$, to the calculations will create  even-narrower resonances. 

A comparison of the top and bottom panels in Figs.~\ref{fig:FBsmallB} and \ref{fig:FBlargeB}  shows that a resonance in  $a(B)$ always corresponds to a threshold bound state with zero binding energy.
Each of these zero-energy bound states can be followed back to a bound state at zero magnetic field, where
a resonance that occurs at larger $B$ has a larger binding energy at $0$~G.
For example, zero-energy bound states that occur around $B=1000$~G have a zero-field
binding energy of $\approx h\times 5$ GHz, outside the range of energies shown in Fig.~\ref{fig:FBlargeB}.
Moreover,  the magnetic moments of the bound states, $-{\rm d}E/{\rm d}B$,
are related to the magnetic moments of closed channels, {\it i.e.} atom-pair channels with energies that are larger 
than that of the entrance channel.  These closed channels have magnetic moments of up to a few times the Bohr 
magneton $\mu_{\rm B}$ with $\mu_{\rm B}/h\approx1.40$ MHz/G relative to that of the entrance channel.

Further analysis of the near-threshold bound state wavefunctions has shown that they originate from coupling among the
last three $s$- and $d$-wave bound states, labeled ${v=-1}$, $-2$, $-3$, respectively, of the $V_{\rm X}(R)$and 
$V_{\rm a}(R)$ potentials. In fact,  bound states with $-0.2\ {\rm GHz}<E/h<0\ {\rm GHz}$ have at least  80\,\%
 of their wavefunctions combined in the $1(0^-)$ and $1(1)$ states.
These observations are consistent with the energy-level density expected from the identical attractive 
long-range $-C_6/R^6$ van-der-Waals tail of the two potentials. Analytical analysis of  bound state energies of a van-der-Waals 
potential by Gao in Ref.~\cite{Gao2000}  shows that for the $C_6$ coefficient of FrAg
the relations $-5.1\ {\rm GHz} \le E_{v=-3}/h \le -1.6\ {\rm GHz}\le E_{v=-2}/h \le -0.23\ {\rm GHz} \le E_{v=-1}/h <0$ GHz hold,
where $E_{v=-3,-2,-1}$ are the energies of the last three bound states.
For a $d$-wave channel the energy intervals satisfy $-7.7\ {\rm GHz} \le E_{v=-3}/h \le -2.8\ {\rm GHz}\le E_{v=-2}/h \le -0.60\ {\rm GHz}\le E_{v=-1}/h <0$ GHz.
Combined with the number of closed  $s$- and $d$-wave channels and their threshold energies this leads to the energy level density seen
in Figs.~\ref{fig:FBsmallB} and \ref{fig:FBlargeB}.
   
Finally, we note that our calculations of our relativistic potentials are not exact. In fact, based on  
electronic-structure calculations using smaller  basis sets,
we conclude that the number of bound states has an uncertainty of at least two and one for  
$V_{\rm X}(R)$ and $V_{\rm a}(R)$, respectively.  This implies that Figs.~\ref{fig:FBsmallB} and \ref{fig:FBlargeB} only show a typical  Feshbach spectrum.  The resonance density in $a(B)$, however, will remain the same for any potential pair as the $C_6$ coefficient for FrAg is sufficiently accurate. In fact, the density is 0.005 G$^{-1}$ for $s$-wave resonances and 0.02 G$^{-1}$ for $d$-wave resonances. The precise locations of Feshbach resonances are unknown. Finally, for  the 
Feshbach spectrum in Figs.~\ref{fig:FBsmallB} and \ref{fig:FBlargeB} the background scattering length away from resonances is negative. Changing the shape of the potentials can lead to a positive value for $a$.
 Reference~\cite{Gribakin1993} showed that for a van-der-Waals potential there is  a 75\,\% chance of a positive scattering length $a$.
Joint experimental and theoretical studies of FrAg are required for  determining the exact locations of magnetic Feshbach resonances. 

\subsection{Formation of ultracold FrAg molecules by STIRAP}

In this subsection, we derive  initial guidelines for the formation of ultracold  ground-state FrAg molecules
by analyzing  transition  dipole moments between the initial, intermediate, and final molecular rovibrational
states in stimulated Raman or STIRAP processes based on the pathway shown in Fig.~\ref{fig:PESs}. 
We can assume that FrAg molecules are first created in a weakly-bound near-threshold $s$-wave vibrational state by a slow
ramp of the magnetic field through one of the $s$-wave Feshbach resonances found in the previous subsection. 
Such ramps are nearly $100\,\%$ efficient \cite{Chin2009}.

For our initial analysis of the stimulated Raman or STIRAP process we make several simplifying assumptions.
First, we do not include the hyperfine and magnetic Zeeman interactions in the description of the weakly-bound 
$s$-wave vibrational states. Based on the realization that the wavefunctions of these bound states have at least 
an 80\,\% character in the a$^3\Sigma^+$ state,  it is reasonable to assume that the Raman process starts
in either the $v=-1$, $-2$, or $-3$ $s$-wave vibrational level of the 1(1) component of the a$^3\Sigma^+$ state.

In the STIRAP-based  formation of ultracold alkali-metal dimers \cite{Ni08,Ye2017}, the intermediate states were 
deeply-bound $v',{J'=1}$ ro-vibrational levels of  $n(\Omega^\sigma)=n(0^+)$ excited states with ${n=2}$ and 3.
We will do so for FrAg as well, but introduce one additional approximation. We ignore non-adiabatic mixing near avoided
crossings between the $2(0^+)$ and $3(0^+)$ states  and focus on the $3(0^+)$ state
as the location of and harmonic frequency near its potential minimum are closer to that of the $1(0^+)$ electronic ground-state potential.
The final state in the STIRAP process is energetically lowest $v=0$, $s$-wave level of the  $1(0^+)$ state.  
(As an aside, note that the electronic dipole moments between the $1(0^-)$ component of the a$^3\Sigma^+$ state and $n(0^+)$ 
states are strictly zero.)

\begin{figure} 
\includegraphics[scale=0.31,trim=0 0 0 15,clip]{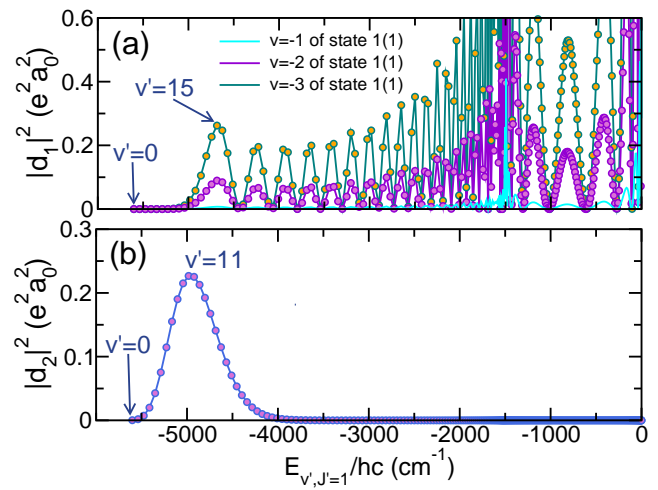}
\caption{Squared vibrationally averaged transition dipole moments of $^{223}$Fr$^{107}$Ag  as  functions of $J'=1$ vibrational binding energies $E_{v'J'}$ of the $3(0^+)$ excited state from  the weakly bound $s$-wave ${v=-1}$, $-2$, or $-3$  
levels of the 1(1) state (panel (a)) and from the most-deeply-bound $s$-wave vibrational level of the $1(0^+)$ electronic ground 
state (panel (b)). 
These two types of transitions correspond to the upward and downward steps of the stimulated Raman process, respectively. 
Filled colored circles correspond bound states $v'$ in the  $3(0^+)$ state. The  $v=-1$ data in panel (a) (cyan colored curve) are
barely visible on the scale of the figure. 
The zero of energy for the $x$ axis is at the Fr(7p$_{3/2}$)+Ag(5s) dissociation limit defined in Fig.~\ref{fig:PESs},
 $ea_0$ is the atomic unit for electric dipole moments, and $e$ is the elementary charge.}
\label{fig:up-down}
\end{figure}

The relevant quantities that are needed to evaluate the effectiveness of the upward and downward  transitions in the 
STIRAP process are the vibrationally averaged dipole moments 
\begin{equation}
d^{(\alpha,\beta)}_{v',v} = \int_0^\infty dR\, \phi^{(\alpha)*}_{v',J'=1}(R) \,d_{\alpha\beta}(R)\, \phi^{(\beta)}_{ v,\ell=0}(R)\, 
\label{VATDM}
\end{equation}
between electronic states ${\alpha=3(0^+)}$ and ${\beta=1(0^+)}$ or $1(1)$.
Here, $d_{\alpha\beta}(R)$ are electronic transition dipole moment shown in Fig.~\ref{fig:TDMs}.  
The  {\it radial} rovibrational wavefunctions   $\phi^{(\alpha)}_{v',J'}(R)$ and $\phi^{(\beta)}_{ v,\ell}(R)$ are unit-normalized 
and  ${v=0}$ for $\beta=1(0^+)$ and ${v=-1}$, $-2$, $-3$ for $\beta=1(1)$.
In principle, Eq.~(\ref{VATDM}) must be multiplied by a dimensionless factor containing the photon polarization 
dependence~\cite{Kotochigova2009}. They are always of the order of one and in view of our other approximation
can be  omitted.

The results of our calculation for the upward  and downward transition dipole moments
as functions of $3(0^+)$ ${J'=1}$ vibrational levels are shown in Figs.~\ref{fig:up-down}(a) and (b), respectively. For the upward transition in panel (a), we 
observe that the dipole moments are on the order of $0.1ea_0$ for many of the vibrational levels $v'$ of the $3(0^+)$ state 
in the bottom half on the potential. For  $3(0^+)$ vibrational levels with energies near the Fr(7p$_{1/2}$)+Ag(5s) and
Fr(7p$_{3/2}$)+Ag(5s) limits and thus with  large, up to $20a_0$, radial extent the dipole moments are significantly larger. That is, 
the overlap of $3(0^+)$ levels with the even-larger extended initial state is largest. Finally, we note that the size of the dipole
moments increase with the binding energy of the initial $s$-wave vibrational state $v$. 
Figure~\ref{fig:up-down}(b) shows the  transition dipole moments for the downward step. Significant transition amplitudes
only occurs for $3(0^+)$ vibrational levels with an energy around $hc\times 5\,000$ cm$^{-1}$ below the 
Fr(7p$_{3/2}$)+Ag(5s) limit.

\begin{figure} 
\includegraphics[scale=0.31,trim=0 0 0 0,clip]{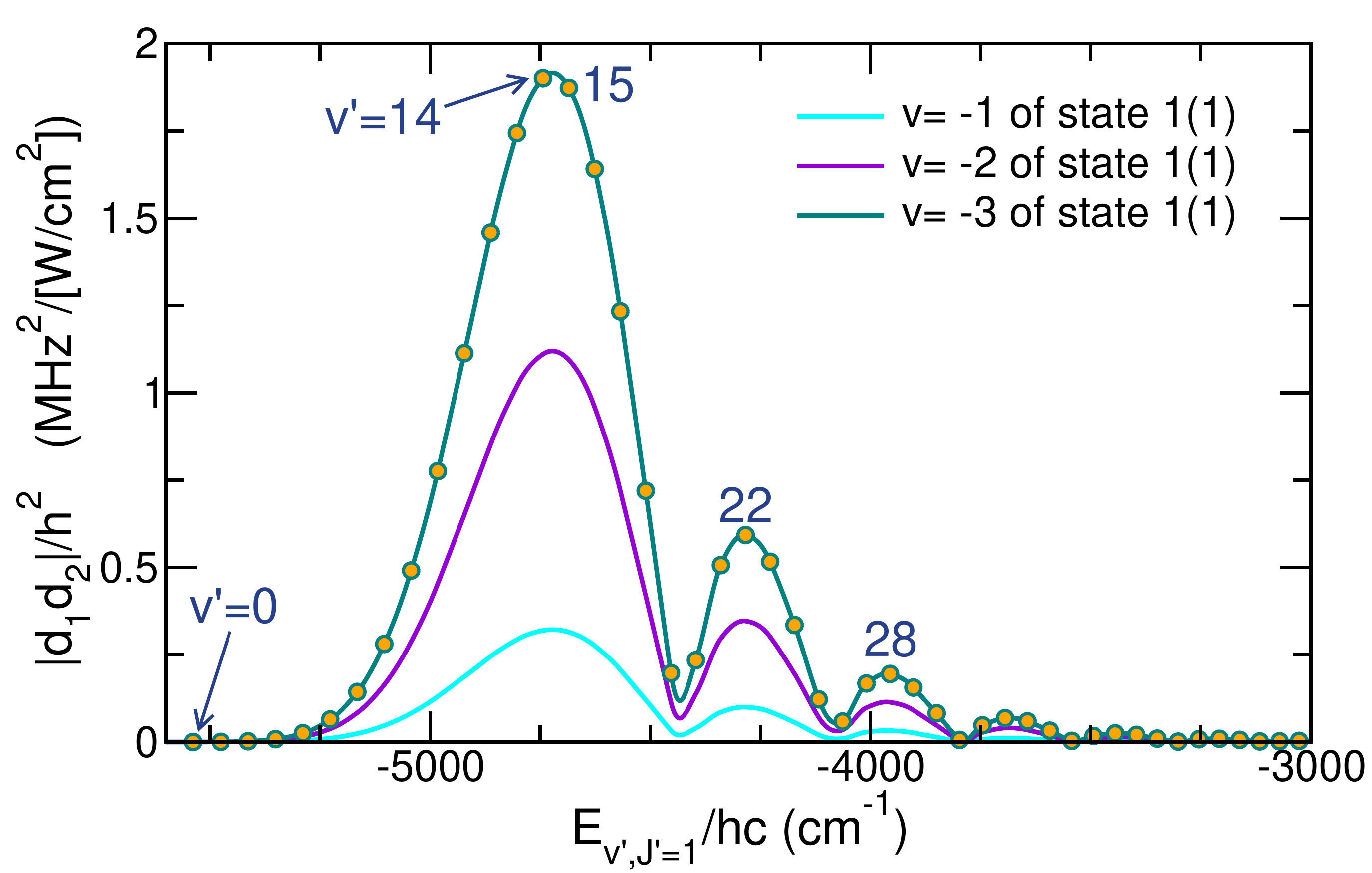}
\caption{Vibrationally averaged two-photon  transition dipole moments of $^{223}$Fr$^{107}$Ag for the stimulated Raman transition from weakly bound $s$-wave $1(1)$ vibrational levels ${v=-1}$, $-2$, and $-3$ to the most-deeply-bound $s$-wave 
vibrational level of the $1(0^+)$ electronic ground state
as functions of $J'=1$ vibrational binding energies $E_{v'J'=1}$ of the intermediate $3(0^+)$ electronic state. 
The zero of energy for the $x$ axis is at the Fr(7p$_{3/2}$)+Ag(5s) dissociation limit defined in Fig.~\ref{fig:PESs}.
The data are derived from Fig.~\ref{fig:up-down}.}
\label{fig:VATDM}
\end{figure}

The  transition amplitude for resonant two-photon, two color Raman transitions is the proportional to 
$(d_1 {\cal E}_1)(d_2{\cal E}_2)/(\Delta_{v',J'=1}+i\gamma_{v',J'=1})$, 
where $d_i$ and ${\cal E}_i$ with ${i=1}$ or 2 are the vibrationally averaged   dipole moments and 
electric field strengths of the lasers for the upward and downward transitions, respectively.
The 
frequencies $\Delta_{v',J'=1}$ and $\gamma_{v',J'=1}$ are the detuning from and the linewidth of rovibrational level $v',J'=1$ of 
the intermediate $3(0^+)$ state, respectively. Figure~\ref{fig:VATDM} shows $d_1d_2$ for the last three $1(1)$ $s$-wave bound 
states as  functions of vibrational energies of the $3(0^+)$ potential. We  see that the best candidates for intermediate state 
are vibrational levels ${v'=14}$ and ${v'=15}$, about $hc\times 1600$ cm$^{-1}$ above the minimum energy of the $3(0^+)$ 
potential. Starting from the $v=-3$ vibrational level of the $1(0^+)$leads to the largest two-photon rates.

\section{Conclusion}

Molecules with unstable isotopes often contain heavy and deformed nuclei and thus possess a high sensitivity to various 
parity-violating effects. In this paper, we theoretically studied the molecular properties of $^{223}$Fr$^{107}$Ag, a molecule 
with exceptional promise in quantum sensing and precision measurements of parity-violating effects. Experimental efforts 
will likely use molecules formed or associated from ultracold laser-cooled Fr and Ag atoms. 
We therefore  determined  adiabatic relativistic electronic energies of ground and excited molecular states as well
as electronic transition dipole moments between them
and  showed that it is feasible to create $^{223}$Fr$^{107}$Ag molecules by two-color photo-association or STIRAP
to its energetically lowest rotational, vibrational state from ultracold $^{223}$Fr and $^{107}$Ag atoms. 

To reach this conclusion, we set up a hyperfine- and Zeeman-resolved quantum coupled-channels scattering calculation
for one microKelvin ground-state $^{223}$Fr and $^{107}$Ag atoms. From these calculations, we showed that
many magnetic Feshbach resonances exist as a function of  applied magnetic field up to 1500 G. 
We estimated that the nearest-neighbor level density of these resonances  is 0.005 G$^{-1}$ for $s$-wave resonances 
and 0.02 G$^{-1}$ for $d$-wave resonances. We also found that the resonances are due to mixing of the last three, most weakly 
bound vibrational  levels of the  $1(0^+)$  and $1(0^-,1)$ potentials. The accuracy of these potentials, however, 
is insufficient to predict the number of molecular bound states and thus of the exact location of Feshbach resonances.
Joint experimental and theoretical efforts are required to determine these quantities.

Secondly, we computed rovibrationally averaged one- and two-photon transition dipole moments
from one of the weakly bound $1(1)$ $s$-wave vibrational levels to the $v=0,$ $s$-wave rovibrational level 
of the $1(0^+)$  ground electronic state. We chose vibrational levels of the adiabatic $3(0^+)$ state as intermediate levels
and suggest that vibrational levels about $hc\times 1600$ cm$^{-1}$ above the minimum energy of the $3(0^+)$ 
potential are the most favorable for FrAg formation.
This suggestion also implies the need for very different laser frequencies for the STIRAP process. 

In the future we hope to improve the quality of our predictions of the stimulated Raman and STIRAP transition strengths.
In this article, we made several approximations to find initial order of magnitude estimates. The most problematic one might 
be the adiabatic approximation of the intermediate $3(0^+)$ state. Non-adiabatic mixing near avoided crossings between 
the $2(0^+)$ and $3(0^+)$ states can be important.

\section{Acknowledgments}

Work at Temple University is supported by the U.S. Air Force Office of Scientific 
Research Grants No. FA9550-21-1-0153 and No. FA9550-19-1-0272, the NSF Grant No. PHY-1908634. 
We thank Dr. D. DeMille, Dr. T. Fleig, and Dr. A. Petrov for fruitful discussions.

\appendix

\section {Electronic structure computations} \label{App1}

We have performed Kramers unrestricted relativistic coupled-cluster calculations with single, double, and perturbative triplet 
excitations (CCSD(T))~\cite{Visscher:1996} using DIRAC program \cite{Dirac21} for the ground $n(\Omega^\sigma)=1(0^+)$ and  $1(0^-)$ states of FrAg corresponding to the Hund's case (a) singlet X$^1\Sigma^+$ state and 
the energetically lower of the two relativistic components of the triplet a$^3\Sigma^+$  state, respectively. The small-core relativistic effective core potential, 
designed for the aug-cc-pwCV5Z-PP basis sets, from Ref.~\cite{Dolg:2012} has been used. In particular, we use the 
ECP78MDF and ECP28MDF core potentials for Fr and Ag, respectively. Reference molecular orbitals and determinants are obtained from relativistic Dirac-Coulomb Hartree-Fock calculations and only  electrons in the outermost  $6s^26p^67s^1$ shells of Fr and $4s^24p^64d^{10}5s^1$  shells of Ag are correlated in the calculations. 

We find that the $1(0^+)$ ground state is well described by a single determinant near the repulsive wall and 
global minimum up to interatomic separations of ${\cal R}_{\rm X}=11 a_0$.  For separations  between $13a_0$ and $14 a_0$ the ground state 
energy has an unphysical maximum. Here, the $1(0^+)$ potential is closest to that of the $2(0^+)$ or A$^1\Sigma^+$ state 
and its electronic wavefunction is multi-reference in nature.  Consequently, the $1(0^+)$ potential can only be used for 
$R\le{\cal R}_{\rm X}$ and the molecular interaction energies must be obtained by subtracting the ground-state monomer energies 
of Fr and Ag obtained with the same Kramers unrestricted CCSD(T) method and basis sets. 
DIRAC calculations of the energies of the $1(0^-)$ state do not suffer from unphysical maxima
and we are able use the results for $R$ up to ${\cal R}_{\rm a}= 19a_0$.
We again subtract the ground-state monomer energies of Fr and Ag to determine  the potential  $V_{1(0^-)}(R)$. 

Potential energies of other electronic states have been calculated within the Generalized Active Space (GAS) approach of 
relativistic four-component  all-electron LUCIA calculations. 
Reference orbitals or spinors have been obtained from open-shell Dirac-Coulomb Hartree Fock calculations with two open 
shell orbitals, namely the 7s orbital of Fr and the 5s orbital of Ag.  The remaining less-extended orbitals are kept 
doubly occupied.  In the end the GAS  approach has 58 inactive  and 38 active 
spinor orbitals. Virtual unoccupied orbitals are build up from the atomic basis set.

Our  choice of GAS allows for single excitations from the 6p shell of Fr,  single excitations from the 4d shell of Ag, 
two excitations from the 6p7s shell pair of  Fr,  as well as two excitations from the 4d5s shell pair of Ag. 
To avoid so-called accidental root flipping, we request  convergence of 10 roots or eigenstates for each $\Omega$.  

\begin{figure}
\includegraphics[scale=0.31,trim=10 0 0 0,clip]{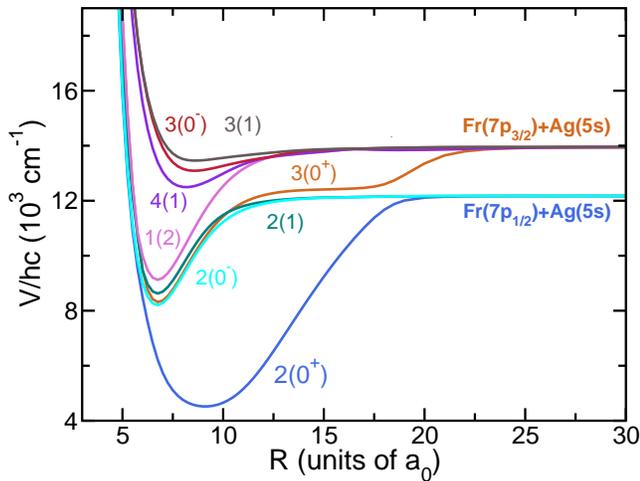}
\caption{Some relativistic electronic excited-state potentials of FrAg as functions of  separation $R$.
Potentials are identified by Hund's case (c) state labels $n(\Omega^\sigma)$ and by
atomic labels for large $R$. The zero of energy 
is at the dissociation limit or threshold of the ground electronic state. }
\label{fig:addPESs}
\end{figure}

The  LUCIA calculations have  been used to determine both potentials and $R$-dependent transition dipole moments.
All $\Omega=0^+$, $0^-$, 1 and 2 potentials dissociating to either the excited Fr(7p$_{1/2}$) or Fr(7p$_{3/2}$) limits
while  Ag remains in its ground state are shown in Fig.~\ref{fig:addPESs}. In the main part of this paper a subset of these potentials, those relevant for STIRAP-based formation of the FrAg molecule,  as well as relevant transition dipole moments  have already been shown.

For the  coupled-channels calculations we need as input potentials $V_{\rm X}(R)$ and $V_{\rm a}(R)$ for
the Hund's case (a) non-relativistic singlet X$^1\Sigma^+$ and triplet a$^3\Sigma^+$ states, respectively.
We can use $V_{\rm X}(R)\equiv V_{1(0^+)}(R)$ for the X$^1\Sigma^+$ state from the CCSD(T) calculations.
For potential of the triplet a$^3\Sigma^+$ state, we must combine the data from the coupled-cluster and LUCIA calculations.
Our CCSD(T) data are more accurate than those from LUCIA calculations. On the other hand coupled-cluster 
calculations and their extensions could not be used to determine the $1(1)$ component of the a$^3\Sigma^+$ state.
Instead we  construct a  $V_{1(1)}(R)$ potential by adding the small positive splitting $V_{1(1)}(R)-V_{1(0^-)}(R)$ 
between the $1(1)$ and $1(0^-)$ states from the LUCIA calculations to the CCSD(T) $V_{1(0^-)}(R)$ potential
It is worth noting that  the small splitting is due to second-order spin-orbit effects with distant 
excited electronic states.
Finally, we use that $V_{\rm a}(R)\equiv( V_{1(0^-)}(R)+2 V_{1(1)}(R))/3$, a weighted
mean or barycenter of the potentials for the two components of the a$^3\Sigma^+$ state,
based on an effective dipolar rank-2 spin-spin Hamiltonian between the electron spins of each of the atoms.

We realize that for separations, where the electron wavefunctions of the atoms barely overlap, {\it i.e.} 
$R>{{\cal R}_{\rm disp}\approx22a_0}$,
both  $V_{\rm X}(R)$ and $V_{\rm a}(R)$ approach the dispersion potential $V_{\rm disp}(R)=-C_6/R^6-C_8/R^8$ 
omitting smaller contributions. The van-der-Waals dispersion coefficient $C_6=1116E_{\rm h}a_0^6$ was already computed in Ref.~\cite{Tomza2021}.  Currently, no value for the $C_8$ dispersion coefficient 
is available. We chose $C_8= 746\,685 E_{\rm h}a_0^8$  inline with typical values for alkali-metal dimers and
leading to a reasonable connection to the DIRAC results for the short-range potentials. 
We then connect each short-range DIRAC potential to the long-range dispersion potential  using extrapolations 
of $V_{\rm X,a}(R)$ and $V_{\rm disp}(R)$ to intermediate-range where ${\cal R}_i<R<{\cal R}_{\rm disp}$,
$V^{\rm tot}_i(R)= [1-s(R;\vec{p})] V_{i}(R) + s(R;\vec{p})V_{\rm disp}(R)$ for $i=$X and a and
step-like functions $s(R;\vec{p})$ with values between 0 and 1 for increasing $R$. Here, $\vec p$ represents 
state-dependent adjustable parameters and  $s(R;\vec{p})$ is based on the trigonometric function $\tanh(x)$.
We have verified that with this procedure $V^{\rm tot}_{\rm X}(R)$ and $V^{\rm tot}_{\rm a}(R)$ do not cross.

\bibliography{FrAg_References}

\end{document}